\documentclass[12pt]{article}

\usepackage{amssymb,amsmath}

\usepackage{graphicx}

\newcommand{\be}{\begin{equation}}
\newcommand{\ee}{\end{equation}}
\newcommand{\ep}{\varepsilon}
\newcommand{\bt}{\beta}

\begin{document}

\begin{center}

{\Large{\bf Statistical Outliers and Dragon-Kings \\as Bose-Condensed Droplets} \\[5mm]

V.I. Yukalov$^{1,2}$ and D. Sornette$^{1,3}$} \\ [3mm]

{\it
$^1$Department of Management, Technology and Economics, \\
ETH Z\"urich, Swiss Federal Institute of Technology, \\
Z\"urich CH-8032, Switzerland \\ [3mm]

$^2$Bogolubov Laboratory of Theoretical Physics, \\
Joint Institute for Nuclear Research, Dubna 141980, Russia \\ [3mm]

$^3$Swiss Finance Institute, c/o University of Geneva, \\
40 blvd. Du Pont d'Arve, CH 1211 Geneva 4, Switzerland}

\end{center}

\vskip 2cm

\begin{abstract}

A theory of exceptional extreme events, characterized by their abnormal sizes 
compared with the rest of the distribution, is presented. Such outliers,
called ``dragon-kings'', have been reported in the distribution of financial 
drawdowns, city-size distributions (e.g., Paris in France and London in the UK), 
in material failure, epileptic seizure intensities, and other systems. Within 
our theory, the large outliers are interpreted as droplets of Bose-Einstein 
condensate: the appearance of outliers is a natural consequence of the occurrence
of Bose-Einstein condensation controlled by the relative degree of attraction, 
or utility, of the largest entities. For large populations, Zipf's law is 
recovered (except for the dragon-king outliers). The theory thus provides 
a parsimonious description of the possible coexistence of a power law 
distribution of event sizes (Zipf's law) and dragon-king outliers. 

\end{abstract}

\vskip 1cm



\newpage

\section{Exogenous versus endogenous causes of outliers}

Outliers in statistical observations are those data that appear to be 
markedly deviating from other members of the statistical sample in which 
they occur [1-3]. There are numerous examples of outliers that, 
generally, can be of two kinds. One type of outliers are those that 
are caused by different errors. For instance, a physical apparatus for 
taking measurements may have suffered a transient malfunction. There 
may have been an error in data transmission or transcription. Outliers 
can arise due to human errors or instrument errors. A sample may have 
been contaminated with elements from outside the population being 
examined. Such erroneous outliers, due to exogenous reasons, are not 
of interest, except that it is the duty of the scientist to recognize
them and remove them for a sound analysis. These outliers are not of 
our concern in the present paper.

Another cause of outliers can be endogenous, due to natural deviations 
in populations. It is these natural outliers whose appearance has to be
understood and explained. Such outliers may seem to be in contradiction 
with the assumed theory, calling for further investigation. 

For example, a number of statistical data are known to be in good 
agreement with power-law distributions. Such power-law type distributions 
have appeared in different branches of statistical analysis of data [4-8] 
and are now referred to as the Pareto law or the Zipf law, depending
on the context and the value of the exponent. Numerous illustrations of 
power laws in a variety of applications can be found in the literature 
dealing with natural languages, information theory, city populations, 
web access data, internet traffic, bibliometry, finance and business,
ecology, biology, genomics, earthquakes, and so on. It is impossible to 
list all that literature, so we shall cite a few works, where further 
references can be found [9-32]. 

In many cases, there are however marked deviations from the power laws,
exhibiting large natural outliers. The number of such outliers is not 
high, usually there is just a single outlier outside the main sample. 
As examples, we can mention the rank-size distributions of French cities, 
where Paris is an outlier, of Great Britain cities, where London is an 
outlier, of Brazilian cities, where S\~{a}o Paulo is an outlier, the 
distribution of Hungarian cities, where Budapest is an outlier, also the 
rank-size distribution of billionaires in certain countries 
(the ``king'' effect), and so on [33]. More examples can be found in 
Ref. [34], where these endogenous outliers are named ``dragon-kings". 

We stress that these endogenous dragon-kings are fundamentally different 
from the concept of ``black swans'' [35], as explained in [34]. The concept 
of black swan is essentially the same as Knightian uncertainty, i.e., a
risk that is a priori unknown, unknowable, immeasurable, not possible to 
calculate. Nassim Taleb thinks of a black swan as an unpredictable extreme 
event of enormous impact, especially in the social sphere (private communication, 
February 2011). One possible incarnation of a black swan is a tail event of 
a power law distribution. In contrast, the dragon-king concept [34] stresses 
the fact that many extreme events are distinguishable by their sizes or by 
other properties from the rest of the statistical population. Dragon-kings 
are argued to result from mechanisms that are different, or that are amplified 
by the cumulative effect of reinforcing positive feedbacks. As a consequence 
of the amplifying mechanism that is specific to the dragon-king appearance, 
they may actually be knowable, and they may be characterized by specific 
precursors. Dragon-kings thus carry ambivalence: (i) on the one hand, they 
occur more often than predicted by the extrapolation of the statistical
distribution calibrated on their smaller siblings; (ii) on the other hand, 
they may be forecasted probabilistically, more than the other large events 
in the tail of the standard statistical distribution.

In order to capture this essential difference with ``black swans'', the term
``dragon-king'' was chosen as follows. First, in some countries, the king and 
his family own or control a large part of the whole country wealth while, at 
the same time, the rest of the population wealth is Pareto distributed. This 
constitutes an example of the coexistence of a power law distribution of 
wealths and of a singular point, the king's wealth that is outside and beyond 
the distribution of the rest of the population. We refer to this as the king 
effect. The term ``dragon'' describes an animal, yes, but an animal of mystical 
and supernatural powers. Hence, similarly to the king, there is the coexistence 
of some properties of the dragon that are common with the rest of the animals 
(wings, tail, claws, etc), together with absolutely abnormal characteristics 
that make the dragon apart from the rest of the animal kingdom.

This debate illustrates that the knowledge of extreme events is still very poor.
The causes for such endogenous outliers are not always clear. Although, we may 
mention that in the theory of phase transitions there are several so-called 
{\it droplet models} exhibiting the appearance of large critical clusters,
essentially overpassing the sizes of all other droplets [36-41]. The occurrence 
of such extreme clusters can happen in different types of interacting statistical 
systems, such as condensed matter [36-41], gravitating matter [42], quark-hadron 
matter [43,44], as well as in social phenomena, e.g., in the clustering of citizen
into cities [45]. 

Some authors connect the existence of the phenomenon of extreme-cluster 
formation with Bose-Einstein condensation. This phenomenon is well known in 
physics and intensively studied both theoretically and experimentally, as can 
be inferred from the recent reviews [46-53]. The possibility of mapping 
different other effects to the Bose-Einstein condensation has also been 
discussed. Among these effects, it is possible to mention the functioning 
of memory [54-56], traffic jams [57], wealth distribution [58], network 
evolution [59], and ecological dominance [60]. 

In the present paper, we advance a new approach for treating statistical 
data of complex systems. The appearance of large natural outliers is 
explained as due to Bose-Einstein condensation. The approach is rather 
simple and general and can be applied to different data samples.

\section{Definitions and assumptions of model}

\subsection{Qualitative analogy between Bose-Einstein condensation of atoms 
and emergence of an outlier in the distribution of city sizes}

Before addressing the mathematical basis of the approach we suggest, let us
give the intuition for why the Bose-Einstein condensation can be connected 
to the problem of characterizing statistical outliers. For this purpose, let 
us recall in a few words what is the essence of the phenomenon of Bose-Einstein 
condensation. In the present context, the most convenient setup is the 
condensation of trapped atoms, for which the energy spectrum is discrete,
contrary to the uniform case, where the spectrum is continuous. 

Suppose that a system of many atoms is confined in a trapping potential, 
with a discrete energy spectrum characterized by the level energies 
$\varepsilon_1, \varepsilon_2, \ldots,  \varepsilon_n, \ldots$. Atoms are 
distributed over these energy levels, so as to achieve the minimal free energy 
for the system. At sufficiently low temperature, a great number of atoms pile 
down in the lowest energy level, as shown in Fig. 1. This concentration of atoms 
in the lowest level is the Bose-Einstein condensation. While the lowest level 
corresponds to the minimal energy, not all atoms are able to occupy it, and there 
are always atoms on other higher energy levels. This corresponds to the condensate 
depletion that is caused by two reasons: temperature fluctuations and repulsive 
atomic interactions. In the presence of the Bose condensate, the distribution of 
atoms over energy levels cannot be given by a single function, but the condensate 
has to be separated out from the distribution of the rest of the atoms. 
Consequently, the {\it condensate is nothing but a statistical outlier}.

A similar phenomenon occurs in a statistical system of agents that can occupy 
different``levels'' corresponding to different ranks, such as the inhabitants of 
different cities. And each given country is finite, similarly to a finite system 
of trapped atoms. Discrete energy levels are analogous to separate cities. People 
tend to live in those cities that are the most convenient and profitable for them. 
Usually, the best opportunities are provided by the largest cities classified by 
the lowest ranks, within a rank-ordering classification. Thus, the largest city 
can become an outlier, if a Bose-Einstein condensation occurs. And similarly to the 
case of atoms, not all people can gather in a single city because of various 
disturbing factors and individual competitive interactions. The concentration of 
people in the largest city is equivalent to the Bose condensation. Then, the 
distribution of inhabitants over all cities can describe all the cities at the 
exception of the city-outlier that has to be separated from the distribution, in 
the same way as it is done for a Bose-condensed system of atoms.

The Bose-Einstein condensation of atoms results from the Bose statistics that 
is at the origin of correlations between atom occupancies over the population of
available energy levels. Analogously, the correlations among people are realized 
by the exchange of information. The choice of cities and the evaluation of their suitability, or fitness, or attractiveness, is always done on the basis of 
the information available to decision makers. The information exchange between 
people plays the role of quantum correlations for atoms.
        
Information processing, of course, requires time and can depend on the 
geometric location of participants, similarly to the processes of atomic 
interactions. The time scaled involved in information exchange is much shorter 
than the typical lifetime of a country. This is in analogy with the smallness 
of the atomic interaction time as compared with the lifetime of a trapped system. Therefore, it is admissible to invoke an equilibrium description, where the short
time-dependent processes are averaged out. As a result, such an equilibrium 
description does not depend on time and on the spatial location of participants.     

In order to summarize the analogies between the Bose-Einstein condensation of 
trapped atoms and the condensation of inhabitants in cities, we give below a 
dictionary connecting these phenomena.

\vskip 0.3cm
\begin{center}

\begin{tabular}{|c|c|} \hline

confining potential     &   {\it country territory} \\ \hline

trapped atoms           &   {\it country citizens} \\ \hline

energy levels           &   {\it separate cities}  \\ \hline     

level number            &   {\it city rank}     \\ \hline  

level distribution      &   {\it rank distribution} \\ \hline
   
quantum correlations    &   {\it information exchange} \\ \hline

thermal fluctuations    &   {\it disturbing factors} \\ \hline

repulsive interactions  &   {\it competitive interactions} \\ \hline

energy minimization     &   {\it utility maximization} \\ \hline

most profitable level   &   {\it most profitable city} \\ \hline
\end{tabular}
\end{center}

\subsection{Definitions and formulation of the model}

Our approach is general, being applicable to various statistical 
data. For the sake of concreteness, we illustrate it for the case 
of the {\it rank-size distribution} of cities in a country. Equally, 
the approach is valid for the {\it rank-frequency distribution} of words 
in a text, as well as for other statistical data in sociology, linguistics, 
economics, and so on.

We consider the situation when the process of city formation has reached 
a stationary regime inside the given country [61]. This condition holds 
by construction in the case of the word-frequency distribution in a given 
text, which is written and therefore fixed.

Let $N$ be the total number of persons (or households or other atomic groups) 
representing the total population of a country. Or this could be the total 
number of words in a text. This means that the population can be grouped 
into {\it characteristic population} elements indexed by $n = 1, 2, \ldots, N$. 
Or this could be the number of word groups, each group consisting of the same 
word. The total number $N$ is assumed to be very large, $N \gg 1$. This 
population of $N$ persons is distributed among $C$ cities. 

The {\it city rank} $\varepsilon(n)$ of a city with $n$ inhabitants 
is defined as the number of cities whose population is larger than or equal 
to $n$ inhabitants. This means that the rank is related to the cumulative 
distribution of population over cities. The ranks are arranged in the 
ascending order, so that the rank of a larger city is smaller:
\be
\label{1}
 \ep(n_1) < \ep(n_2) \qquad (n_1 > n_2 ) \;  .
\ee
Rank $1$ corresponds to the largest city, rank $2$ to the second largest city,
and so on. Our aim is to find a relation between the city rank and its 
characteristic population.

The relation and our derivation of the Bose-Einstein condensation phenomenon
in the distribution of cities relies on three key ingredients.

\vskip 0.3cm
{\bf Assumption 1}. {\it The characteristic feature of an inhabitant selecting 
a city can be described by the concept of {\it utility factor} $w(\varepsilon)$, 
which is assumed to be a function only of the city rank $\varepsilon$}.

\vskip 0.3cm
{\bf Justification}. Each city has evolved during its history, in competition
and through diverse complex interactions with other cities in the country.
The characteristic feature of an inhabitant selecting a city can be described 
by the factor $w(\varepsilon)$ characterizing the stationarity state of the 
relative attraction, according to the usefulness for the decision maker, of 
each of the $C$ cities to the diverse individuals in the total population of 
$N$ persons. The utility factor $w(\varepsilon)$ takes the values within the 
interval $[0, 1]$; the larger its numerical value, the more convenient  and the 
more attractive is the city. The factor $w(\varepsilon)$ is a decreasing function 
of rank $\varepsilon$.

\vskip 0.3cm
{\bf Assumption 2}. {\it The attraction factor $w(\varepsilon)$ is taken to be 
an exponentially decreasing function of rank}:
\be
\label{7}
 w(\ep) = b e^{-\bt\ep} \qquad ( b > 0 \; , \; \bt > 0 ) \;  .
\ee

\vskip 0.3cm
{\bf Justification}. This is a natural choice, very often assumed in the 
biological literature dealing with fitness [62-71]. The parameter $\bt$ is called 
the decline parameter. The fitness factor decreases with rank, capturing the 
fact that a smaller city (higher rank) is less attractive, in general, due to less 
opportunities for job, cultural entertainments, synergies and so on [72]. The 
attraction of large cities is usually associated with increasing returns to scale 
and economies of scale [73]. The utility function in decision theory is also often 
taken in the exponential form [74,75].

\vskip 0.3cm
{\bf Assumption 3}.  {\it The probability that a city of rank $\varepsilon$ has 
a population of not less than $n$ inhabitants is assumed to be a multiplicative 
function of the utility factors for each person}:
\be
\label{2}
 p_n(\ep) = a w^n(\ep) \; .
\ee

\vskip 0.3cm
{\bf Justification}. This expression (\ref{2}) derives from the assumption that separate 
individuals make their choices independently from each other. As the overall 
quality or attraction of a city for a given person is completely captured by 
the utility factor $w(\varepsilon)$, the population of a given city then results 
from independent choices performed by each inhabitant, which is equivalent to
taking the product of the factors $w(\varepsilon)$.

\vskip 0.3cm

\section{Derivation of the rank-size cumulative distribution}

The probability $p_n(\ep)$ given by (\ref{2}) has to be normalized as
\be
\label{3}
 \sum_{n=1}^N p_n(\ep) = 1 \;  ,
\ee
in order to express that each city is certainly inhabited. This normalization, 
taking into account that
$$
\sum_{n=1}^N w^n \simeq \frac{w}{1-w} \qquad ( N \gg 1 ) \; ,
$$
yields
$$
a = \frac{1-w}{w} \;   .
$$
Therefore, the probability  $p_n(\ep)$ defined by (\ref{2}) acquires the form
\be
\label{4}
p_n(\ep) = [ 1 - w(\ep) ] w^{n-1}(\ep) \;   .
\ee
 
The characteristic population of a city of rank $\varepsilon$ is defined 
as the expectation value
\be
\label{5}
 n(\ep) = \sum_{n=1}^N n p_n(\ep) \;  .
\ee
This, in view of the equality
$$
\sum_{n=1}^N n w^n \simeq \frac{w}{(1-w)^2} \qquad (N \gg 1) \;   ,
$$
gives the expression
\be
\label{6}
 n(\ep) = \frac{1}{1-w(\ep)} \;  .
\ee
Expression (\ref{6}) is valid for any type of attraction factor $w(\varepsilon)$. 
Generally, the latter could be taken in different forms, but here we employ the
expression (\ref{7}). 

With the form (\ref{7}) of the attraction factor, the characteristic population 
(\ref{6}) becomes
\be
\label{8}
 n(\ep) = \frac{e^{-\bt\ep}}{e^{\bt\ep} - b} \;  .
\ee
Introducing the notation
\be
\label{9}
 \mu \equiv \frac{1}{\bt} \; \ln b  
\ee
reduces Eq. (\ref{8}) to 
\be
\label{10}
 n(\ep) = \frac{e^{\bt(\ep-\mu)}}{e^{\bt(\ep-\mu)}-1} \; .
\ee
This has the form of the typical Bose-Einstein function describing the 
population distribution, with the nominator playing the role of a 
degeneracy factor, the rank $\varepsilon$ playing the role of energy, 
and $\mu$ playing the role of a chemical potential.

Inverting Eq. (\ref{10}) gives the rank of a given city
\be
\label{11}
\ep(n) = \mu + \frac{1}{\bt} \; \ln \left ( \frac{n}{n-1} \right )
\ee
as a function of its characteristic population size $n$. A priori, the 
rank cannot be smaller than one for all population sizes $n$ larger than one:
\be
\label{12}
 \ep(n) \geq 1 \qquad ( n > 1 ) \;  .
\ee   

In this way, Eq. (\ref{10}) gives the expectation value of the population size
for a city of rank $\varepsilon$. Here the rank is fixed. Equation. (\ref{11}),
conversely, defines the rank for the given characteristic population size.
Recall that $n(\varepsilon)$ is a cumulative distribution, hence its sum over
the ranks does not define the total country population.

The formal definition of a city is a disputed and complex issue, that we 
do not address here. Let us just mention that cities have more than one 
inhabitant, being relatively large and permanent settlements, with 
administrative, legal, or historical status based on local law. Thus, there 
should be in general one more restriction on the rank, when the lowest 
characteristic population is fixed by some number $m$, so that all $C$ cities 
have populations not smaller than this minimal number $m$. Then the boundary 
condition follows:
\be
\label{13}
\ep(m) = C \;   ,
\ee
which defines the chemical potential
\be
\label{14}
 \mu = C - \; \frac{1}{\bt} \; \ln \left ( \frac{m}{m-1} 
\right ) \;  .
\ee

Expression (\ref{11}) retrieves a variant of Zipf's law for the largest cities. 
Indeed, taking $n \gg 1$ allows one to expand the logarithm and obtain
\be
\label{11bis}
\ep(n) = \mu + \frac{1}{\bt} \cdot  \frac{1}{n}  \qquad (n \gg 1) \; ,
\ee
relating the city rank to the inverse of the population size. The only 
difference with the standard formulation of Zipf's law is the term $\mu$ that 
gives a small correction. Since $n(\varepsilon)$, by definition, is positive, 
then $\mu$ should be smaller than the minimal $\varepsilon$ equal to one. 
Hence $\mu < 1$.

\section{Bose-Einstein condensation into dragon-king cities}

It turns out that the distributions (\ref{10}) and (\ref{11bis}) are not the 
whole story. Indeed, the boundary condition (\ref{13}) may sometimes 
disagree with the population distribution (\ref{10}). The occurrence 
of such a disagreement signifies the appearance of an anomaly, that we 
term a ``dragon-king'' city [34], which is  equivalent to Bose-Einstein 
condensation. 

To demonstrate how this happens, let us consider the population of the most 
inhabited city of rank one
\be
\label{15}
 N_1 \equiv n(1) = \frac{e^{\bt(1-\mu)} }{e^{\bt(1-\mu)} -1} \;  .
\ee
By its definition, this is a finite positive number, which requires 
that $\mu$ be smaller than one,
\be
\label{16}
 \mu < 1 \qquad ( 0 < N_1 < \infty ) \;  .
\ee
The latter implies that the decline parameter $\beta$ has to be limited by 
the inequality
\be
\label{17}   
 \bt < \bt_c \qquad ( \mu < 1 ) \;  ,
\ee
where the critical value is defined by
\be
\label{18}
 \bt_c \equiv \frac{1}{C-1} \; \ln \left ( \frac{m}{m-1} 
\right ) \;  .
\ee

When condition (\ref{17}) holds true, the boundary condition (\ref{13})
is compatible with the population distribution (\ref{10}). Therefore, 
the rank-size distribution of cities follows formula (\ref{11}) with $\mu$ 
given by equality (\ref{14}). 

However, when $\beta$ becomes larger than the critical value $\beta_c$,  
the boundary condition (\ref{13}) becomes incompatible with the population 
distribution (\ref{10}). A large value of $\beta$ implies a relatively much 
stronger attraction to the first rank compared with the higher ranks. It is 
therefore expected that the largest city, rank $1$, plays a special role. 
Indeed, considering that the largest city of rank $1$ is an outlier of the 
distribution of all other cities, that we refer to as a dragon-king, we should 
exclude it from the statistics described by distribution (\ref{11}). 
The remaining cities continue to be described by this distribution. The 
situation is completely analogous to the Bose-Einstein condensation, where 
the role of the dragon-king is played by the Bose condensate. In that sense, 
the dragon-king represents a condensate droplet, with its inhabitants 
playing the role of condensate particles. 

In general, it is possible that several largest cities could be outliers 
(dragon-kings) of the distribution of city sizes and thus 
should be excluded from the description offered by formula (\ref{11}). In 
physics, this would correspond to the occurrence of granular condensate
consisting of several grains, or droplets, of condensed particles in the
surrounding of uncondensed matter [52,76]. For instance, if $k$ cities 
are dragon-kings, then for the population $n(k)$ to be a positive number, 
$\mu$ has to be smaller than $k$, hence the decline parameter $\beta$ has 
to be constrained by the inequality 
\be
\label{bbt}
\bt < \frac{1}{C-k} \; \ln \left ( \frac{m}{m-1} \right ) \;   .
\ee
In that case, a series of condensation transitions would arise, with 
the condensation of the largest city, then of the second largest one, and 
so on as the decline factor $\beta$ increases through the succession of the 
critical values, corresponding to a larger and larger mismatch between the 
attraction factors of large cities and smaller ones. 

For the convenience of presenting the formulas, let us introduce the parameter 
that can be called the effective {\it temperature}
\be
\label{19}
 T \equiv \frac{\bt_c}{\bt} \;  .
\ee
This parameter quantifies the level of noise causing the dispersion of the country
inhabitants among different cities. Then the rank-size distribution (\ref{11}) 
takes the form
\be
\label{20}
 \ep(n) = \mu + \frac{T}{\bt_c} \; \ln \left ( \frac{n}{n-1}
\right ) \;  ,
\ee
with the chemical potential 
\be
\label{21}
 \mu = C - (C-1) T \;  .
\ee
For high temperature $T > 1$, one has $\mu < 1$, and there is no 
condensation. The value $T_c = 1$ is the critical point of the starting
condensation, where $\mu = 1$. Below this temperature $T < 1$, the largest city 
of rank $1$ falls out of the data sample, becoming a dragon-king, with other cities
remaining uncondensed and described by the rank-size distribution (\ref{11}).

With these notations, the generalized Zipf's law, obtained for large $n$, reads as
\be
\label{22}
 \ep \simeq \mu + \left ( \frac{T}{\bt_c} \right ) \frac{1}{n}
\qquad ( n \gg 1 ~~~ \& ~~~\mu<1)~,
\ee
relating the rank $\ep$ of a given city to its population size $n$.
In this way, our model not only gives the interpretation of outliers in the 
distribution of city sizes as Bose-condensed droplets, but it also provides 
a possible mechanism for the Zipf's law.

\section{Discussion}

We have suggested a novel approach for deriving rank-size distributions. 
First, we obtain the Zipf law that has been well documented for a variety 
of statistical data. Secondly, according to this approach, the appearance 
of outliers of the distributions, that we have referred to as 
``dragon-kings'' [34], is equivalent to the Bose-Einstein condensation. 
The objects of the first ranks, such as the largest cities, when becoming 
outliers, are similar to Bose-condensed droplets. For concreteness, we 
followed the interpretation related to the rank-size distribution of cities. 
But the approach can be applied for interpreting the appearance of outliers 
of other nature, e.g., in the rank-frequency distribution of words 
in different texts.

In physics, the Bose-Einstein condensation is a collective coherent phenomenon. 
The same interpretation applies to our derivation through the three assumptions
underlying our model, which are set to capture the collective organization
of cities over their historical development. Consequently, the appearance of 
dragon-kings can also be understood as the result of collective effects resulting 
in the coherent accumulation of agents in these outliers. The Bose 
condensation is a phase transition. Hence, we propose that the occurrence 
of ``dragon-kings"  [34] is also a kind of a phase transition. Other related 
mechanisms for the formation of dragon-kings are also associated with phase 
transitions. Let us mention generalized percolation transition, where the 
infinite cluster plays the role of the outlier or dragon-king. Another example 
is that of the synchronization between moving objects, like oscillators [77], 
which can give rise to the coexistence of a power law distribution of event 
sizes and of dragon-kings [78-80].

When dealing with statistical data, there are phenomenological recipes that
allow one to suspect the presence of outliers [1-3]. The simplest hint that 
the object of rank $1$ is an outliers is when
\be
n(1) - n(2) \gg n(2) - n(3) \;  .
\ee
In the approach we suggest, the procedure would be as follows. For the given
numbers $m$ (of a minimum city size) and $C$ (number of cities in the country)
characterizing the considered statistical set, one should fit the function 
$\varepsilon(n)$ to the given data, thus, defining the parameter $T$. One 
should compare the fittings with and without the object suspected to be an 
outlier. Comparing these fittings and the related values of $T$ and $\mu$, one
could conclude, in line with the above theory, whether there is condensation 
or not. Condensation should correspond to the low temperature $T < 1$.

The suggested method of describing outliers applies to the sets of given data, 
such as city sizes or word frequencies. One could ask the question whether the
method could be transferred to systems with time evolution, such as stock market
data accompanied by crashes? The principal answer to this question is yes, 
provided the whole set of data is given. Omitting details, the idea for using 
the method to temporal data would be as follows. Let a database of time series 
be given, where one can define the drawdowns, occurring at different times and
quantified by some index. And let $n$ be the value measuring the fall of the 
index from some previous peak value. Suppose $\varepsilon(n)$ is the 
number of drawdowns whose fall index is larger than $n$. This $\varepsilon(n)$
plays the role of the drawdown rank. Then we may follow the general consideration
described above. The largest drawdown becomes an outlier when condensation occurs.
Then this condensed outlier represents a market crash.
 
The aim of the present paper has been the development of a general theory. So, 
here we limit ourselves by the principal points. Applications to particular 
examples involve discussions of technical problems of fitting methods, which 
is out of the scope of the present paper. Different applications will be 
treated in separate publications.

\newpage

\begin{figure}[h]
\centerline{\includegraphics[width=8cm]{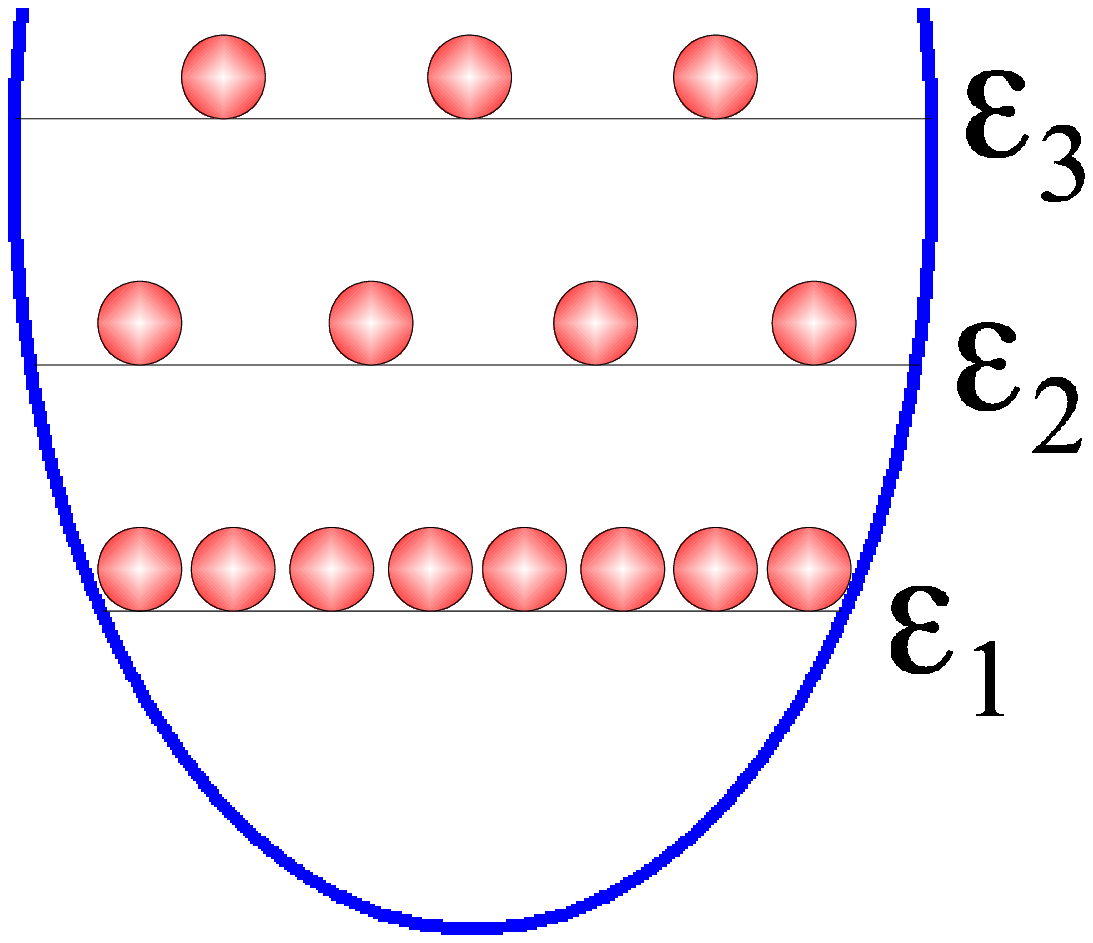}}
\caption{Scheme of the Bose-Einstein condensation of trapped
atoms}
\label{fig:Fig.1}
\end{figure}

\end{document}